\documentclass[twocolumn,prb,showpacs,floatfix,aps]{revtex4} 
\usepackage{amsmath,eucal}
\usepackage{graphicx}
\usepackage{pifont}
\usepackage{verbatim}
\usepackage[latin1]{inputenc}
\usepackage{calc}
\usepackage{subfigure}
\usepackage{rotating}
\begin{document}

\author{J. Prasongkit}
\email{jariyanee.prasongkit@fysik.uu.se}
\affiliation{Division of Materials Theory, Department of Physics, Uppsala University, SE-75121 Uppsala, Sweden.}

\author{A. Grigoriev}
\affiliation{Division of Materials Theory, Department of Physics, Uppsala University, SE-75121 Uppsala, Sweden.}

\author{G. Wendin}
\affiliation{Department of Microtechnology and Nanoscience-MC2, Chalmers University of Technology, SE-41296 G$\ddot{o}$teborg, Sweden.}

\author{R. Ahuja} 
\affiliation{Division of Materials Theory, Department of Physics, Uppsala University, SE-75121 Uppsala, Sweden.} 
\altaffiliation{Applied Material physics, Department of Materials and Engineering, Royal Institute of Technology (KTH), 10044, Stockholm, Sweden}

\title{Interference effects in phtalocyanine controlled by H-H tautomerization: a potential two-terminal unimolecular electronic switch}

\date{\today}

\begin{abstract}
We investigate the electrical transport properties of two hydrogen tautomer configurations of phthalocyanine (H$_2$Pc) connected to cumulene and gold leads. Hydrogen tautomerization affects the electronic state of H$_2$Pc by switching the character of molecular orbitals with the same symmetry close to the Fermi level. The near degeneracy between the HOMO and HOMO-1 leads to pronounced interference effects, causing a large change in current for the two tautomer configuratons, especially in the low-bias regime. Two types of planar junctions are considered: cumulene-H$_2$Pc-cumulene  and gold-H$_2$Pc-gold. Both demonstrate prominent difference in molecular conductance between ON and OFF states. In addition, junctions with gold leads show pronounced negative differential resistance (NDR) at high bias voltage, as well as weak NDR at intermediate bias.
\end{abstract}

\pacs{
85.65.+h, % Molecular electronic devices
73.63.-b, % Electronic transport in nanoscale materials and structures
31.15.Ne, % Electronic structure of atoms and molecules: theory (Self-consistent-field methods)
03.65.Yz % Decoherence; open systems; quantum statistical methods
}

\maketitle

\section{Introduction}
The ability to control current in molecular junctions is a prerequisite for functional devices in molecular electronics. Of particular interest for nanoelectronics are two-terminal switches characterized by two distinct current-voltage characteristics (IVC) representing ON- and OFF-states, switchable by exceeding threshold voltages in hysteretic junctions. With such two-terminal components one can in principle build nanoscale memory and resistor-diode logic for scalable architectures \cite{Snider2005a,Snider2005b}. 

A major challenge is to develop two-terminal unimolecular switches (UMS) where the switching is truly intramolecular, and not due to cooperative or interface effects in multi-molecule junctions. One UMS example is provided by a recent experiment by L\"ortscher {\em et al.}, using a mechanically controlled break junction (MCBJ) to demonstrate switching and hysteretic current-voltage characteristics (IVC) of a single Tour-wire molecule \cite{Lortscher2006,Lortscher2007}. 

An UMS should be electrically coupled and immobilized in the circuit, yet have enough freedom to perform switching. Up to now, several different kinds of molecular switches have been proposed, most of them involving a change of molecular conformation \cite{Donhauser2003,Qiu2004,Choi2006}, or redox state  \cite{Gittins2000}. An interesting case related to the present discussion is an SnPc switch \cite{Wang2009} where the current is determined by the vertical position of the tin atom, controlled by a scanning tunneling microscope (STM).

Recently, Liljeroth \textit{et al.}  \cite{Liljeroth2007} proposed a new type of molecular switch where the switching is mediated by a tautomerization reaction of metal-free naphthalocyanine. Using a low-temperature STM, a voltage pulse at the STM tip can induce a change in the orientation of the hydrogen atom-pair at the center of naphthalocyanine, leading to switching between low and high conductance. The fact that naphthalocyanine does not change position or conformation of the molecular framework could be very beneficial for controlling molecular electronic circuits. If the molecule can be integrated in the circuit, it will not move or change its outer shape upon switching.
 
There is now a variety of techniques that can be used for contacting and investigating single molecules
(see e.g. Refs.~\cite{Tao2006,Tao2007,Chen2007,Takayanagi1998,MolenLiljeroth2010,Osorio2010,Danilov2008,Kervennic2009,Grill2010,Bombis2009,Bombis2010} ).
However,  when it comes to interconnecting single molecules to functional electronic circuits, there are currently no working solutions. The circuits will necessarily have to be based on various forms of self-assembly of building blocks \cite{Grill2007,MothPoulBjorn2009}, but the field is wide open, and extensive fundamental research will be needed before useful functional systems can emerge. 

One possibility to wire circuits could be to first deposit the molecules and then grow the wires. This case has recently been investigated by Okawa \textit{et al.} \cite{Okawa2001,Takajo2007,Okawa2010}, with the pronounced goal to fabricate a single-molecule integrated circuit. Nano-clusters of phthalocyanine molecules are adsorbed on a molecular layer of diacetylene, and chain polymerization is initiated by applying a pulsed bias voltage to the row of diacetylene molecules to which a phthalocyanine pentamer is adsorbed. In this way Okawa \textit{et al.} \cite{Okawa2010} recently succeeded in connecting two polydiacetylene chains to the same pentamer phthalocyanine molecule.

In this paper, we focus on the planar electrical transport properties of a UMS based on hydrogen tautomerization in metal-free phthalocyanine (H$_2$Pc) connected to semi-infinite metallic wires of carbon (cumulene)  \cite{Prasongkit2010} or gold. The transport properties are examined as a function of the electronic structure of the tautomer state of the molecule, the type of leads, and the coupling between electrodes and the molecule. Our results show that switching the orientation  of the H-H atom pair in the H$_2$Pc cavity can significantly change the conductance of the molecule, effectively leading to ON- and OFF-states for both cumulene (Fig. \ref{IVC_cumulene}) and gold (Fig. \ref{IVC_Au}) leads. Moreover, in the case of gold we find that negative differential resistance (NDR) appears at high bias voltages due to reduced transmission through one of the contacts.

A few important  clear limitations should be noted from the start: we treat neither (i) the influence of any substrate nor (ii) the switching mechanism and stability (usefulness) of the switch. 
Liljeroth \textit{et al.} \cite{Liljeroth2007} have demonstrated experimentally that the switch works in principle in vertical transport (STM) with the molecule lying flat on an insulating monolayer on top of a metallic substrate. 
In our work we instead probe theoretically the horizontal (planar) transport properties of a "freestanding" 2D system of  a H$_2$Pc molecule plus leads. Connecting molecular devices in the 2D plane will be essential for building functional circuits - here we present some first results  for planar transport of some simple model systems in order to assess their potential usefulness as switches or non-linear elements.

The mechanism of bias-driven tautomerization has recently been attributed to electron-phonon coupling\cite{Fu2009,Sarhan2009}, implying that the switching can most probably be triggered by either an STM-like current pulse into the molecular plane, or a current pulse along the H$_2$Pc plane. The results of Liljeroth \textit{et al.}  \cite{Liljeroth2007} showed the STM-driven molecule to be bistable, however fluctuating (telegraph type of noise) between the two conductance levels with time (and especially under elevated bias). We would like to point out that the interaction of the H$_2$Pc with the underlying substrate stabilizes the OFF and ON states by increasing the transition barrier. In the related bistable case of Ag-C$_{60}$-Ag \cite{Danilov2008}, the hysteretic behaviour was more stable, except at high bias where fluctuating switching dominated.

As stated at the beginning, a switch has two properties: (i) an IVC with two branches, ON and OFF; and (ii) a mechanism for switching between those two branches.  The OFF state can typically be generated by (a) interrupting a single conduction path by disrupting wave function overlap; or (b) by interference of two conduction paths, placing an interference minimum at the Fermi level.

In the present calculation, the difference between the ON and OFF states is produced by a two-channel interference effect, involving nearly degenerate HOMO and HOMO-1 orbitals and controlled by the H-H tautomer switch. In a molecular context, similar interference effects were discussed early on by Sautet and Joachim \cite{Sautet1988,Sautet1989} for transmission through benzene described as two-path interferometer, and has recently been elaborated for benzene and other systems in several papers using state of the art methods \cite{Cardamone2006,Ke2008,Darau2009, thygen10, thygen10a}. In all of these cases, the basic idea is to influence the interference by varying the phase shift along one of the paths, controlling the transmission at the Fermi level and implementing a transistor or switch.

In the present paper we study the case where the two interfering states remain pinned at the edge of the bias window. Therefore the entire IVC is controlled by interference: the zero-bias destructive interference minimum in the OFF-state stays within the voltage window, following the Fermi level of one of the leads, while the transmission peaks stay outside. In the absence of the other transmission peaks (conducting states) in the voltage window this makes it possible, in principle, to follow e.g. the OFF-state with increasing bias until the molecule switches to the ON-state, generating a hysteretic IVC \cite{Liljeroth2007}. Other methods for switching would be via current pulses \cite{Liljeroth2007} or photoexcitation \cite{Sautet1988,Sautet1989}.

\section{Computational method}

To model the H$_2$Pc junctions, in principle the external leads are coupled directly to the outer rim of the H$_2$Pc  molecule. However, the computational method requires that part of the leads are treated as an extension of the molecule, forming an extended molecule and scattering region coupled to semi-infinite leads. For coupling the semi-infinite cumulene leads to the H$_2$Pc  molecule (Fig. \ref{IVC_cumulene}) we attach two cumulene-like spacers -CH=C=C=C via a $\pi $-bond, expected to provide a stable and transparent coupling. For coupling semi-infinite gold wires to H$_2$Pc (Fig. \ref{IVC_Au}), we model this situation by using short atomic Au chains, Au$_{\mbox{{\scriptsize n}}}$ (n=1 and 2), connected to Au(001) leads, as well as one case of semi-infinite Au chains.

H$_{2}$Pc is connected to two different kinds of electrodes, carbon and gold chains. Such two-probe systems are divided into three regions: left and right electrodes and scattering region. The scattering region contains a portion of semi-infinite electrodes to account for electronic and geometric relaxation at the metal-molecule interface. The geometrical setup procedure of two electrode models and computational details can be briefly described as follows:

(i)	cumulene-H$_{2}$Pc-cumulene

First, the free H$_{2}$Pc molecule was optimized. Two isomers (a) H-H pair horizontal (H$_{2}$Pc$_{\mbox{{\scriptsize h}}}$) (b) H-H pair vertical (H$_{2}$Pc$_{\mbox{{\scriptsize v}}}$) were considered. To setup the cumulene chain electrode system, one hydrogen atom per contact site of optimized H$_{2}$Pc$_{\mbox{{\scriptsize h}}}$ and H$_{2}$Pc$_{\mbox{{\scriptsize v}}}$ was removed and substituted with a (-CH=C=C=C) unit on each side, characterized as a scattering region of the extended molecule. The substituent unit was linked to H$_{2}$Pc at the peripheral group. All configurations, all atoms of scattering region were relaxed. The scattering region was sandwiched between semi-infinite carbon linear chains, cumulenes, consisting of carbon double bonds. Optimum molecule-lead distances were obtained via energy minimization.

(ii) Au-H$_{2}$Pc-Au

Possible adsorption configurations of the scattering region were selected to closely resemble the experimental arrangement of Nazin {\em et al.} \cite{Nazin2003}, where both short atomic gold chains and the molecule were assembled on a metallic NiAl(110) substrate. In our model, the gold atoms were placed at $\eta^{2}$ sites of opposite benzene rings of H$_{2}$Pc. We considered different models for the gold leads: (a) Au(001) semi-infinite needle leads connected via short gold chain, labelled as H$_{2}$Pc$_{\mbox{{\scriptsize h}}}$/H$_{2}$Pc$_{\mbox{{\scriptsize v}}}$@Au{\mbox{{\scriptsize n}}}-Au(001), n=1 and 2, (b) semi-infinite gold chain lead coupled to the H$_{2}$Pc molecule, labeled as  H$_{2}$Pc$_{\mbox{{\scriptsize h}}}$/H$_{2}$Pc$_{\mbox{{\scriptsize v}}}$@Au-chain.  For all systems, the Au-Au positions and bond length of the gold atomic chain in both scattering region and the leads were fixed at 2.88 {\AA} \cite{Nazin2003}. The short Au chains were connected to the Au(001) electrodes at the hollow site.

 The geometry optimization of all systems above has been implemented in the localized-atomic-orbital code SIESTA package \cite{Soler2002}. The GGA-PBE \cite{Perdew1996} and LDA-PW92 \cite{PerdewWang1992} approximations were used for the exchange correlation functional of cumulenes and gold electrodes, respectively, and real-space integrations was performed using 200 Ry cutoff.  The atomic core electrons were modeled with Troullier-Martins norm-conserving pseudopotential \cite{Troullier1991}, and the valence states were 2s2p for C and N, 1s for H, and 5d6s for Au. In both models, the super cells were created with large enough vacuum regions to prevent interaction due to mirror image. We used a double zeta basis with polarization orbitals (DZP) for H, C, N atoms and a single zeta basis with polarization (SZP) for Au atoms. On each atom, which was allowed to move, was relaxed until the force becomes less than 0.01 eV/ {\AA}. Convergence of total energy with respect to k-points sampling was tested up to 3x3x1 Monkhorst-Pack k-point mesh, yielding 6 k-points in the BZ.

To investigate electrical transport properties of two-probe systems, we performed non-equilibrium Green's function (NEGF) technique based on density functional theory-based theory (DFT) as implemented in the TranSIESTA package \cite{Brandbyge2002}. The exchange-correlation potential was approximated by GGA-PBE and LDA-PW92 for cumulenes and gold electrode, respectively. DZP was used for H, C, N atoms while SZP was used for Au atom. When a bias voltage was applied to the junction, the electrical current was calculated by integrating the transmission function over the bias window. In this work, the current-voltage characteristics have been calculated in the bias region from -2 to 2 V, where the potential bias was incremented in steps of 0.2 V.

In the present calculations, the HOMO-LUMO gap is about 1.2 eV (Figs. \ref{mpsh_cumulene} and \ref{TEV_cumulene}), and the Fermi level lies in the vicinity of the HOMO, determined by self-consistent charge transfer. In the experimental setup of Liljeroth {\em et al.}  \cite{Liljeroth2007} the HOMO-LUMO gap is about 2.5 eV, while the gas phase gap is about 4 eV. 
It is well-known and understood that local approximations of the density functional underestimates the HOMO-LUMO gap, and consequently often overestimates the conductance. Nevertheless, NEGF-DFT calculations with standard local exchange-correlation potentials are widely used and provide benchmark results, and we believe that the present calculations provide good estimates of current-voltage characteristics (IVC), especially the differences induced by switching between the tautomer states. A recent discussion of HOMO-LUMO gaps in CuPc can be found in Ref.\citenum{Marom2008}, and the results of more advanced (and time consuming) calculation using the so-called GW approximation for benzene on graphite (0001) is given in Ref.\citenum{Neaton2006} and for Au-S-benzene-S-Au junctions in Ref.\citenum{Quek2009} and \citenum{thygen}.

%%%%%%%%%%%%%%%%%%%%%%%%%%%%%%%%%%%%

\section{Results and Discussion}
\subsection{Cumulene-H$_{2}$Pc-Cumulene}

\begin{figure*}[tp]
\begin{minipage}[t]{1.0\linewidth}
\begin{center}
\includegraphics[width = 13 cm]{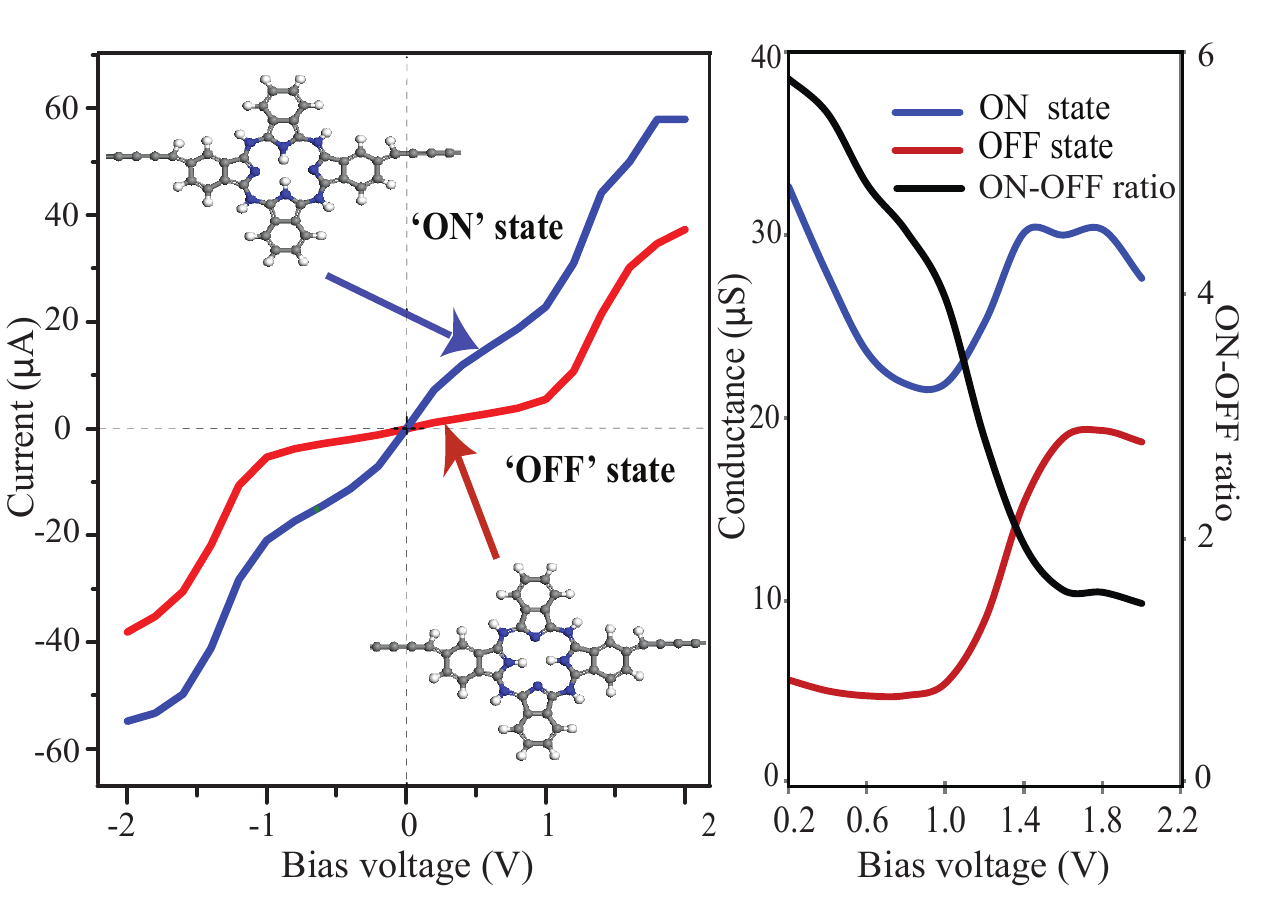}
\end{center}
\end{minipage}
\vspace{-20pt}
\caption{(Color online) The calculated \emph{I-V} characteristics of cumulene(\emph{E}$_{\mbox{{\scriptsize F}}}$+eV/2)-H$_{2}$Pc-cumulene(\emph{E}$_{\mbox{{\scriptsize F}}}$-eV/2) junctions for two cases of inner hydrogen H-H pair orientation in H$_{2}$Pc, obtained in the bias region from -2 to 2 V. Note that the IVC is ohmic - this is not a tunnel junction. The right panel shows the conductance of ON-state and OFF-state and the ON/OFF conductance ratio. For $V>0$, the conductance is calculated as \textit{I}/\textit{V}. The zero-bias conductance is given by the ratio of the ON and OFF transmission coefficients, and takes the value of 431. 
}
\label{IVC_cumulene}
\end{figure*}

Fig. \ref{IVC_cumulene} exhibits the \emph{I}-\emph{V} characteristics (IVC) and corresponding differential conductance for the two cases of hydrogen atoms orientation, H$_{2}$Pc$_{\mbox{{\scriptsize h}}}$ and H$_{2}$Pc$_{\mbox{{\scriptsize v}}}$, in the bias region from -2 to 2 V. The mirror symmetric structure of the H$_{2}$Pc$_{\mbox{{\scriptsize h}}}$ and H$_{2}$Pc$_{\mbox{{\scriptsize v}}}$ molecules results in symmetric \emph{I-V} curves. H$_{2}$Pc$_{\mbox{{\scriptsize v}}}$ performs as the ON-state, and H$_{2}$Pc$_{\mbox{{\scriptsize h}}}$ as the OFF-state.

We first discuss the $T(E,V=0)$ zero-bias part of the transmission spectrum of H$_{2}$Pc$_{\mbox{{\scriptsize h}}}$ and H$_{2}$Pc$_{\mbox{{\scriptsize v}}}$ in Fig. \ref{mpsh_cumulene} and Fig. \ref{TEV_cumulene} in some detail.
The molecular projected self-consistent Hamiltonian (MPSH) eigenvalues (a-b in the OFF and c-d in the ON states, marked as lines in Fig. \ref{mpsh_cumulene}, correspond well to the transmission resonance peaks.

The orientations of the molecular eigen-orbitals depend on the position of the pair of hydrogen atoms. As seen in Fig. \ref{mpsh_cumulene}, the rotation of the H-H pair switches the character of the HOMO and HOMO-1 eigenstates. The HOMO of the OFF-state (a) is weakly coupled to left and right leads and gives rise to a narrow transmission peak close to the Fermi level (\emph{E}=0),
while the HOMO-1 (b) is strongly coupled to the leads, resulting in a broad transmission peak {\em well below} the Fermi level (Fig. \ref{TEV_cumulene}b).

\begin{figure*}[ht]
\begin{minipage}[t]{1.0\linewidth}
\begin{center}
\includegraphics[width = 13 cm]{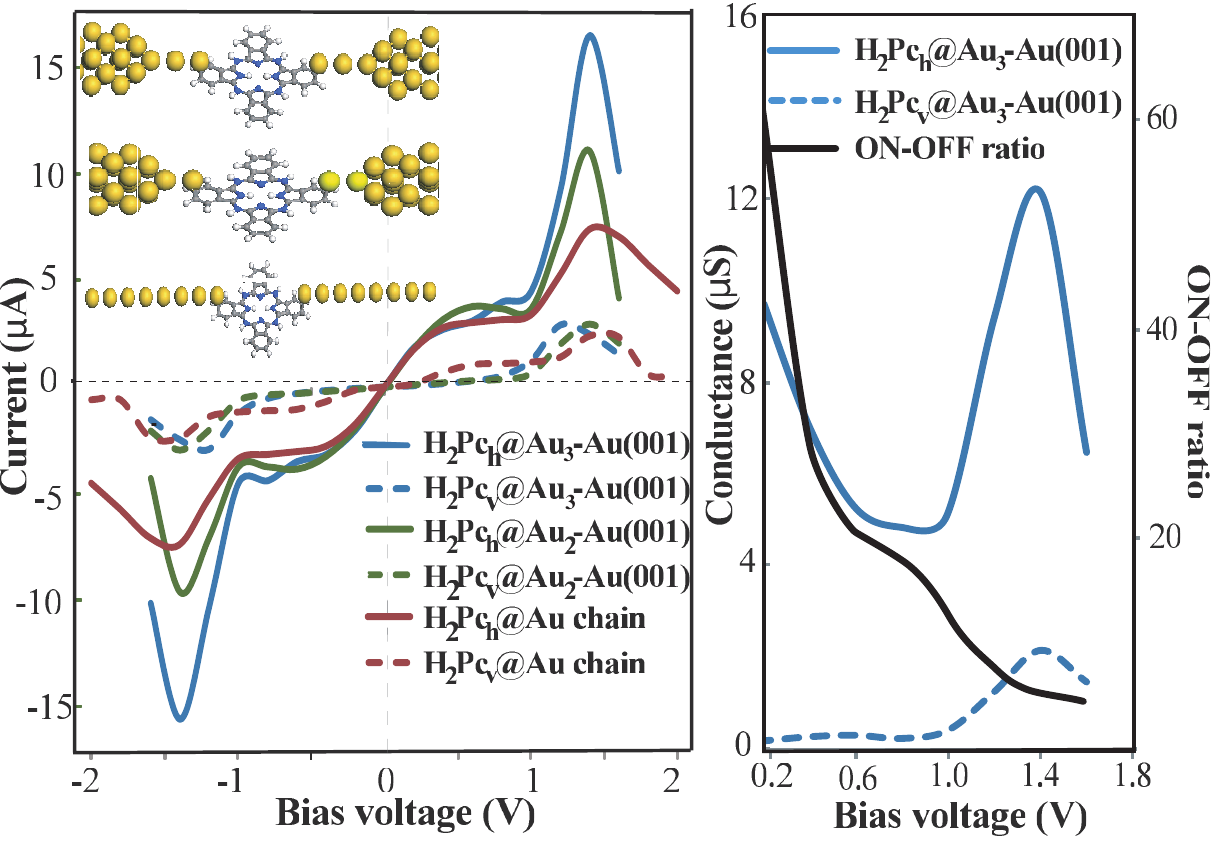}
\end{center}
\end{minipage}
%\vspace{-40pt}
\caption{(Color online) \emph{I}-\emph{V} characteristics for three cases of Au-H$_{2}$Pc-Au junctions for two different H-H pair orientation. Note that the IVC is ohmic - this is not a tunnel junction. The right panel shows the conductance  of H$_2$Pc$_v$@Au$_2$-Au(100) (ON-state) and H$_2$Pc$_h$@Au$_2$-Au(100) (OFF-state), and ON-OFF conductance ratio. For $V>0$, the conductance is calculated as \textit{I}/\textit{V}. The zero-bias conductance is given by the ratio of the ON and OFF transmission coefficients, and takes the value of 86. Also note the appearance of negative differential resistance (NDR) around 1.4 V bias in all cases, as well as weak NDR around 0.7 V in Au(001)-Au$_2$-H$_{2}$Pc-Au$_2$-Au(001).
}
\label{IVC_Au}
\end{figure*}

\begin{figure}[tp]
\begin{minipage}[t]{1.0\linewidth}
\begin{center}
\includegraphics[width = 9 cm]{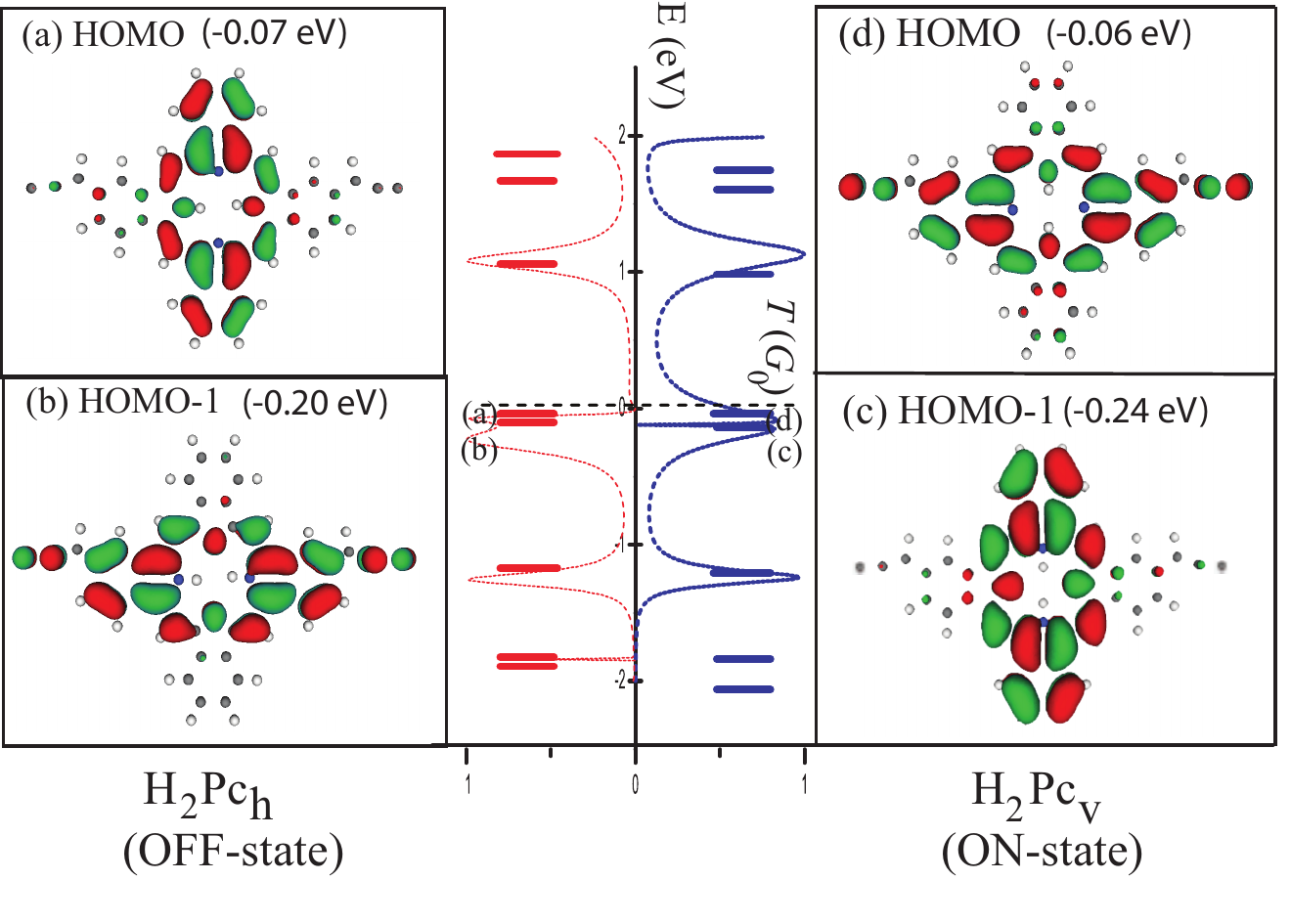}
\end{center}
\end{minipage}
%\vspace{-40pt}
\caption{(Color online) The transmission spectra, the MPSH eigenvalues and the MPSH eigenstates around Fermi level of the 'OFF' and 'ON' configurations of cumulene-H$_{2}$Pc-cumulene.}\label{mpsh_cumulene}
\end{figure}

Turning the H-H pair knob to the ON-state reverses the orbital character of the HOMO and HOMO-1 levels and changes the orbital character at the Fermi level.  Now, the HOMO of the ON-state (d) is strongly coupled to the left and right leads, resulting in a broad transmission peak {\em spanning} the Fermi level (Fig. \ref{TEV_cumulene}a).

The bias dependence of the transmission spectra $T(E,V)$ illustrated in Fig. \ref{TEV_cumulene} explains the basic reason for the large ON/OFF conductance ratio in the 0-1 V bias range: the OFF-state transmission resonance stays almost entirely below the Fermi level of the lead for which chemical potential $\mu$ = \emph{E}$_{\mbox{{\scriptsize F}}}$-\emph{eV}/2 goes down under applied bias, while the ON-state resonance extends into the bias voltage window. Apart from this, the transmission spectra are rather similar.

Since the transmission near the Fermi level proceeds via two channels (orbitals), one can expect interference effects to affect line shapes and widths of the spectral peaks.
Fig. \ref{TEV_cumulene}  indeed provides a nice illustration of Fano resonances and interference in resonance transmission involving a narrow and a broad (continuum) resonance \cite{Fano1961,FanoCooper1968,Wendin1970,Grigoriev2005,Papadopoulos2006}, showing the whole range of lineshapes from a pure window resonance (interference minimum) ( Fig. \ref{TEV_cumulene}(a), ${V}=0$) to strongly asymmetric resonances with pronounced interference minima (Fig. \ref{TEV_cumulene}(a), $V \ge 0.4$) (also seen in the high-lying LUMO peaks).
More importantly, however, the interference has the most pronounced effect on the spectral density at the Fermi level in the OFF-configuration, the Fermi level falling in an interference minimum at zero bias (Fig. \ref{TEV_cumulene}(b), ${V}=0$). 

The H$_2$Pc molecule therefore functions as an interferometer where, by rotating the H-H pair by ninety degrees, we reverse the asymmetry of the line shape of the transmission through the HOMO/HOMO-1 orbitals, placing an interference minimum well below the Fermi level (ON state), or in the HOMO-LUMO gap, suppressing the transmission at the Fermi level (OFF state).

It should be noted that Sautet and Joachim described electronic interference produced by a benzene molecule embedded in a polyacetylene chain \cite{Sautet1988} and applied this to a molecular switch controlled by photoinduced tautomerization \cite{Sautet1989}. The results demonstrated the sensitivity of the zero-bias transmission to the relative position and interaction of the states, disturbed through the tautomerization. Those interferometer effects are similar to the present case. An important difference, however, is the behaviour of such an interferometer under applied finite bias voltage.

\begin{figure}[tpb]
%\begin{minipage}[t]{1.0\linewidth}
\begin{center}
\includegraphics[width = 9 cm]{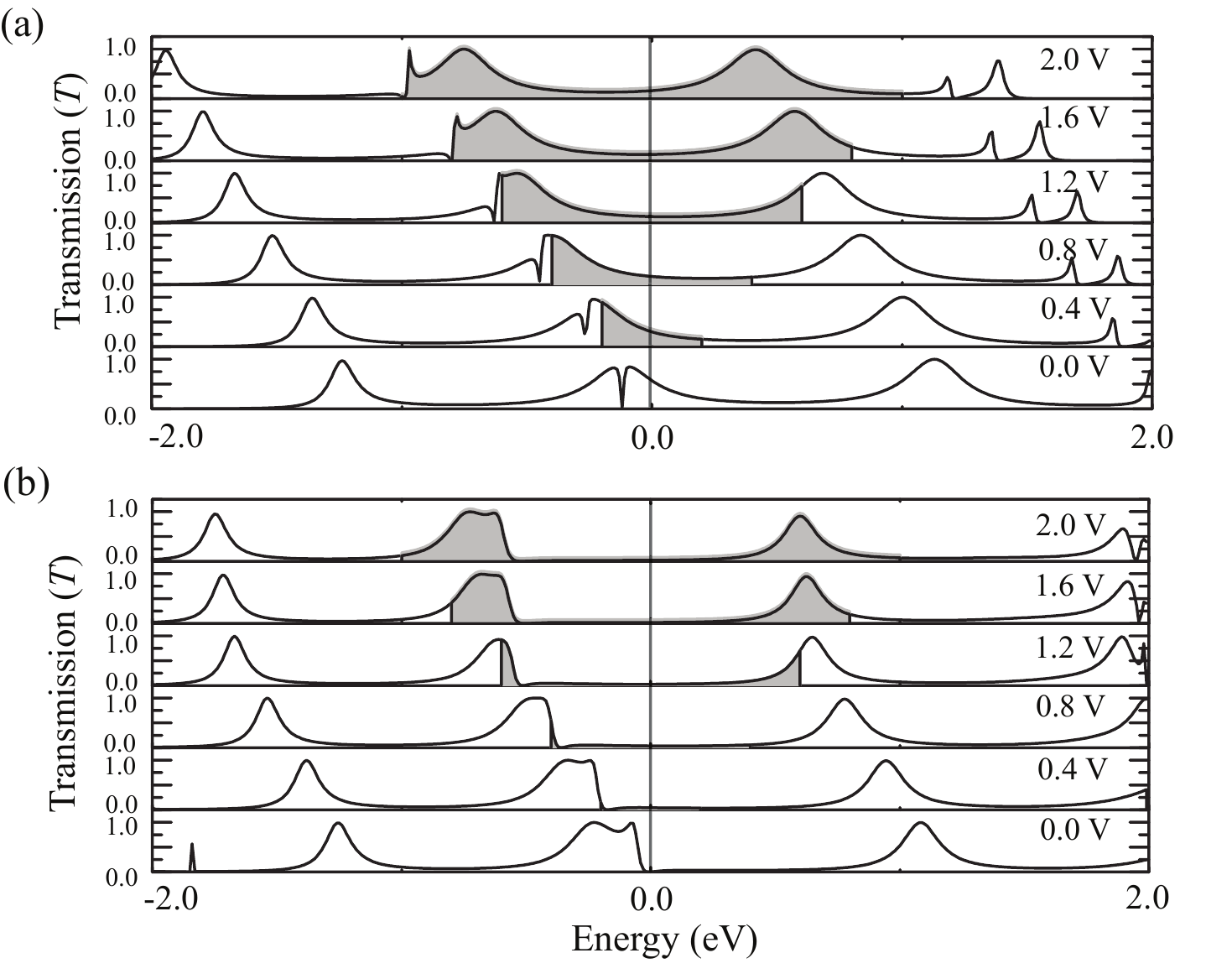}
\end{center}
%\end{minipage}
\vspace{-25pt}
\caption{Bias dependent transmission of cumulene-H$_{2}$Pc-cumulene as a function of energy E. (a) ON-state; (b) OFF-state. The Fermi energy is set to zero; the shaded area indicates the bias voltage window. Notable features are (i) the transmission lines moving down, pinned to the downgoing right lead; (ii) the decoupling from the leads in (b) above 1.2V bias; (iii) the two interfering transmission channel and the Fano lineshapes.
}
\label{TEV_cumulene}
\end{figure}

From the bias voltage dependence of the transmission peaks in Fig. \ref{TEV_cumulene} we can draw a number of conclusions regarding charge transfer, level pinning, and bias voltage drop over the junction.
Starting with the ON-state in Fig. \ref{TEV_cumulene}a, we note that the entire transmission spectrum shifts almost uniformly to lower energy in the direction of the downgoing lead with chemical potential $\mu$ = \emph{E}$_{\mbox{{\scriptsize F}}}$-\emph{eV}/2. At zero bias (\emph{V}=0), charge transfer between H$_2$Pc and both leads places the HOMO resonance at the common Fermi level  (\emph{E}$_{\mbox{{\scriptsize F}}}$=0) of the junction. With increasing bias \emph{V}$>$0, Fig. \ref{TEV_cumulene}a shows that the HOMO-resonance becomes roughly pinned to the lead with downgoing potential, reflecting bias-dependent electron transfer from H$_2$Pc to the lead (positively charging the molecule) to keep the molecular level of H$_2$Pc equal to the chemical potential of the lead, $\mu$ = \emph{E}$_{\mbox{{\scriptsize F}}}$-\emph{eV}/2.

The bias voltage drop in  Fig. \ref{Elstatpot_cumulene}a confirms this picture:  the main bias voltage drop occurs  in H$_2$Pc near the left contact (top panel), and most of the bias voltage increase is distributed over the bulk of the H$_2$Pc molecule, the levels therefore following the chemical potential of the right, downgoing lead.

Examining the ON state on Fig. \ref{TEV_cumulene}a, one notes that the broad HOMO, the HOMO-2 and LUMO move according to \emph{E}$\sim$ -0.4V, not \emph{E}$\sim$ -0.5V, while the narrow HOMO-1 and the high-lying LUMOs move as \emph{E}$\sim$ -0.5V. This reflects the localization of the corrsponding orbitals: e.g. the HOMO-1(Fig. \ref{mpsh_cumulene}(c)) is well localized in a region controlled by the right lead (Fig. \ref{Elstatpot_cumulene}), while the HOMO (Fig. \ref{mpsh_cumulene}(d)) extends into a region with upgoing potential $\mu$ = \emph{E}$_{\mbox{{\scriptsize F}}}$+\emph{eV}/2 controlled by the left lead.

Now considering the OFF-state in Fig. \ref{TEV_cumulene}b, the motion of spectral peaks is the same as for the ON-state in the 0-1V bias voltage region. However, above 1V the levels begin to float, becoming independent of bias voltage. This signals that there is no longer any net charge transfer from H$_2$Pc to the right lead. Indeed, as seen in Fig. \ref{Elstatpot_cumulene}b, the electrostatic potential does not change much over the central-left part of the molecule in the  $1.6\le V\le2.0$ interval.

A closer look at Fig. \ref{Elstatpot_cumulene}a,b reveals some interesting similarities and differences in the voltage drop and screening between the ON- and OFF-states. In the ON-state (Fig. \ref{Elstatpot_cumulene}a), the bias voltage drop is quite stable, but it nevertheless shifts from the center to the left edge of the 5-ring pyrrole group in the $0\le V\le2$ bias interval.  In the OFF-state (Fig. \ref{Elstatpot_cumulene}b), the bias voltage drop is first located around the left edge of the 5-ring pyrrole group in the $0\le V\le1$ bias interval, and then jumps towards the center of the molecule in the $1.2\le V\le1.6$ bias interval (and in fact shifts back a little in the $1.6\le V\le2$ range).

In the OFF-case the conduction state is localized over the central part of the H$_2$Pc molecule (essentially Fig. \ref{mpsh_cumulene}(top, left)) and the voltage drop moves from the left edge to the center of the charge distribution. It is then understandable that the spectral peaks first follow the right lead and then become bias-independent.

In the ON-case, on the other hand, the conduction state is delocalized over the H$_2$Pc molecule and the leads (Fig. \ref{mpsh_cumulene}(top, right)), and the bias voltage drop basically stays put at the center of the left part of the charge distribution. It is then understandable that the spectral peaks follow the right lead over the entire bias region.

The HOMO coupled to the electrodes in the ON state provides an open system with good conductance and metallic-like screening. It is charge-transfer connected to the right lead and (de)charges to stay close to the chemical potential of the right lead. This picture does not change in the investigated $0\le V\le2$ range.

On the other hand, the HOMO in the OFF state is a more closed system with less transparent coupling to contacts. As a consequence, the charge transfer to the right lead will be more difficult to maintain with increasing bias voltage. This means that the molecule will charge up (accept electrons) and float up in energy. This will shift the bias voltage drop toward the right electrode, as already discussed in connection with Fig. \ref{Elstatpot_cumulene}b. These conclusions are, of course, dependent on the nature of the electrodes used to wire the molecule. To be able to identify the effect of the molecular-electrode coupling, we therefore now also consider gold electrodes.

\subsection{Au-H$_{2}$Pc-Au}

Fig. \ref{IVC_Au} presents IVCs for the two H$_{2}$Pc$_{\mbox{{\scriptsize h}}}$/H$_2$Pc$_{\mbox{{\scriptsize v}}}$ tautomer systems connected to several types of gold leads:  3-D semi-infinite needles connected via short 1-D atomic Au-chains (the chains arranged as in the experimental setup by Nazin \textit{et al.} \cite{Nazin2003}), as well as 1-D Au wires. In the case of gold leads, the ON and OFF states correspond to the H$_{2}$Pc$_{\mbox{{\scriptsize h}}}$ and H$_{2}$Pc$_{\mbox{{\scriptsize v}}}$ tautomers, respectively.  In comparison with cumulene, this reversal of ON and OFF states, relative to the orientation of the H-H pair, is due to the symmetry of the coupling orbitals at the H$_2$Pc-Au contacts.

\begin{figure}[tp]
%\begin{minipage}[t]{1.0\linewidth}
\begin{center}
\includegraphics[width = 8 cm]{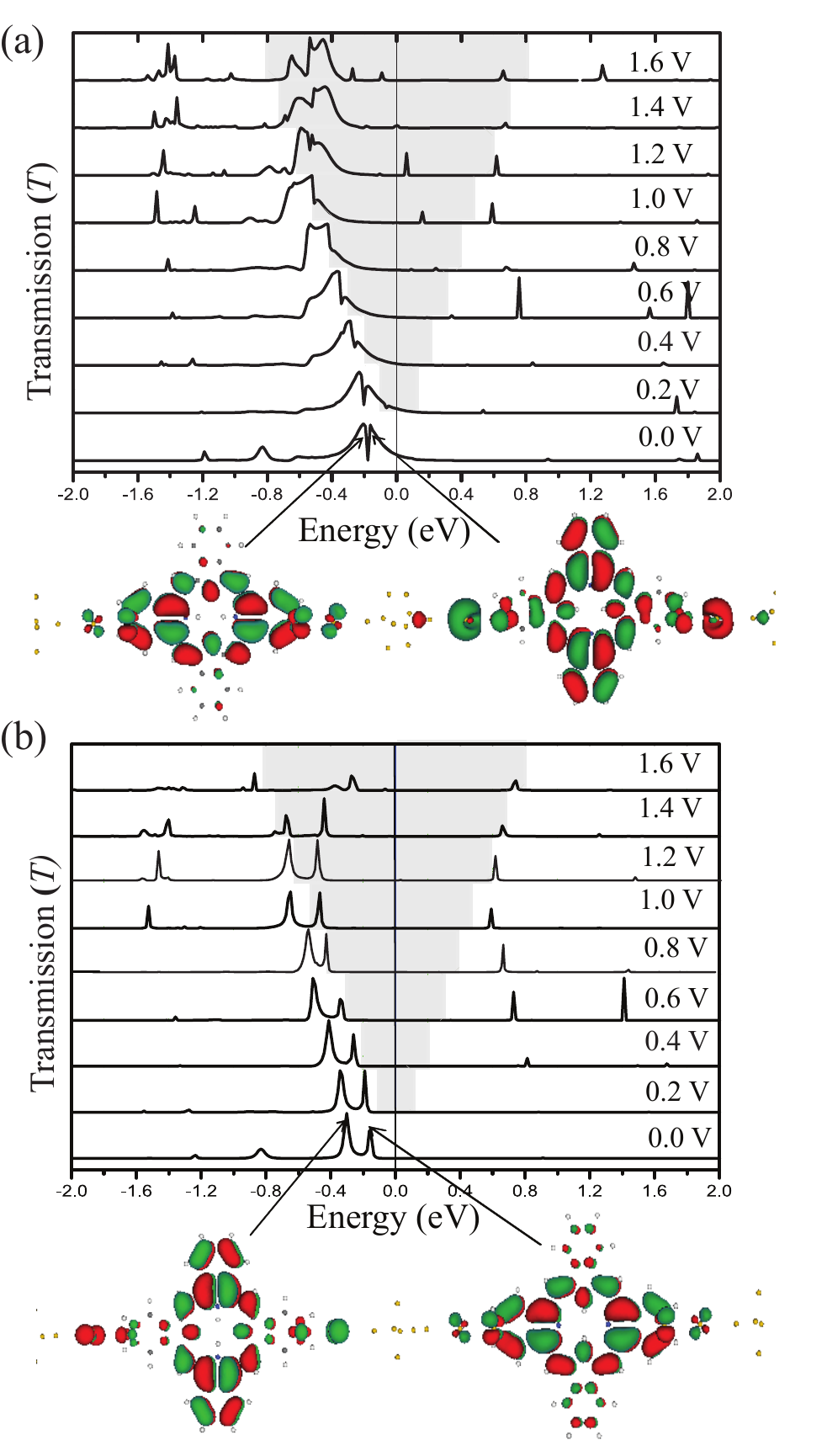}
\end{center}
%\end{minipage}
\vspace{-20pt}
\caption{(Color online) Bias dependent transmission of Au(100)-Au$_3$-H$_2$Pc-Au$_3$-Au(100) as a function of energy \emph{E}. The Fermi energy is set to zero; the shaded area indicates the bias voltage window.
(a) Top panel: ON-state; (b) bottom panel: OFF-state.}
\label{TEV_Au}
\end{figure}

The detailed structure of the IVCs in Fig. \ref{IVC_Au}  can be understood from the bias-voltage dependence of the transmission spectra of Au(100)-Au$_3$-H$_2$Pc-Au$_3$-Au(100), as shown in Fig. \ref{TEV_Au}(a) and Fig. \ref{TEV_Au}(b) respectively.
We first discuss the $T(E,V=0)$ zero-bias part of the transmission spectrum in Fig. \ref{TEV_Au} in some detail. The linewidths with Au leads ($\sim0.2eV$) are considerably narrower than with cumulene leads ($\sim0.4eV$), manifesting a weaker coupling to the leads. It should be understood that the molecule is only physisorbed at the electrodes, bridging the gold chains arranged at the surface\cite{Nazin2003}. The hydrogen tautomerization switches the orbital character of the HOMO and HOMO-1 levels. The profiles of the conduction peaks result from the interference between HOMO and HOMO-1, giving rise to one symmetric broad peak with a dip (ON-state, much like in the case of the cumulene electrodes), and two separate peaks with steep edges  (OFF-state). The spectrum around the Fermi level ($-0.5<E<0$) therefore derives from HOMO and HOMO-1, with different detailed orbital character upon hydrogen tautomerization, turning the H-H "knob".

The zero bias MPSH HOMO and HOMO-1 orbitals of two tautomer systems are shown in Fig. \ref{TEV_Au}. Similar to the cumulene wire leads in Fig. \ref{mpsh_cumulene}, the inner-hydrogen tautomerization of H$_2$Pc switches the molecular orbital character of HOMO and HOMO-1. One molecular orbital shows the interaction between the gold $d_{xz}$-orbital and the carbon $p$-orbital (weak overlap), while the other one shows the interaction between the gold $s$-orbital and the carbon $p$-orbital (strong overlap).
Concerning the transmission at the Fermi energy, in the ON-state the electrons can propagate from the left to the right electrode. In the OFF-state, the electrons are almost totally reflected (at the contact to the right lead, as can be shown with transmission eigenchannel analysis), resulting in the conduction being almost blocked.

\begin{figure}[tp]
%\begin{minipage}[t]{1.0\linewidth}
\begin{center}
\includegraphics[width =9 cm]{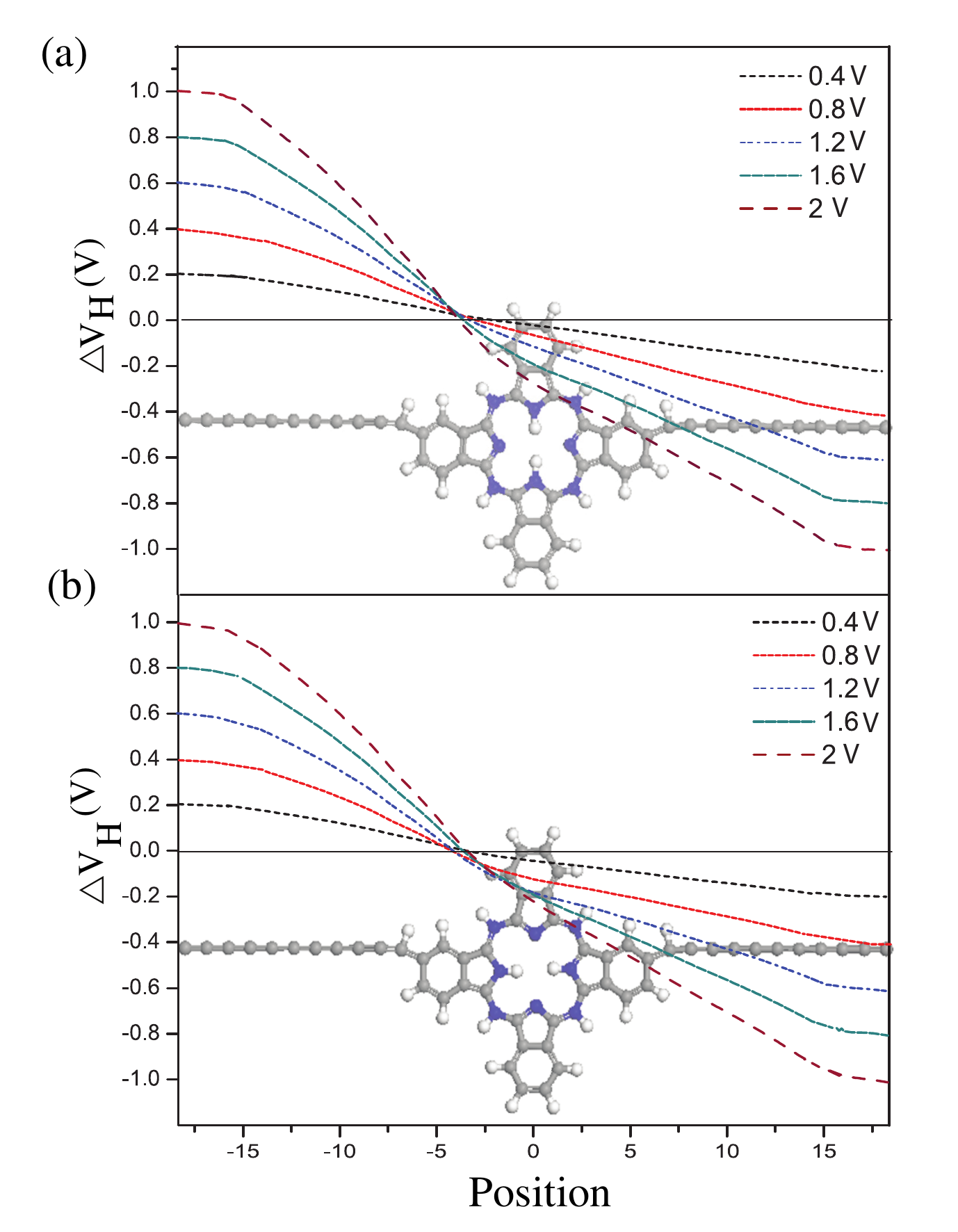}
\end{center}
%\end{minipage}
\vspace{-20pt}
\caption{(Color online) Electrostatic difference (Hartree) potential $\Delta V_H=V_H(V)-V_H(V=0)$, showing the bias-voltage drop over the junction in the (a) ON and (b) OFF states of cumulene-H$_{2}$Pc-cumulene. For negative bias polarity one gets the mirror image of the bias voltage distribution.}
\label{Elstatpot_cumulene}
\end{figure}

\begin{figure}[tp]
\includegraphics[width = 9 cm]{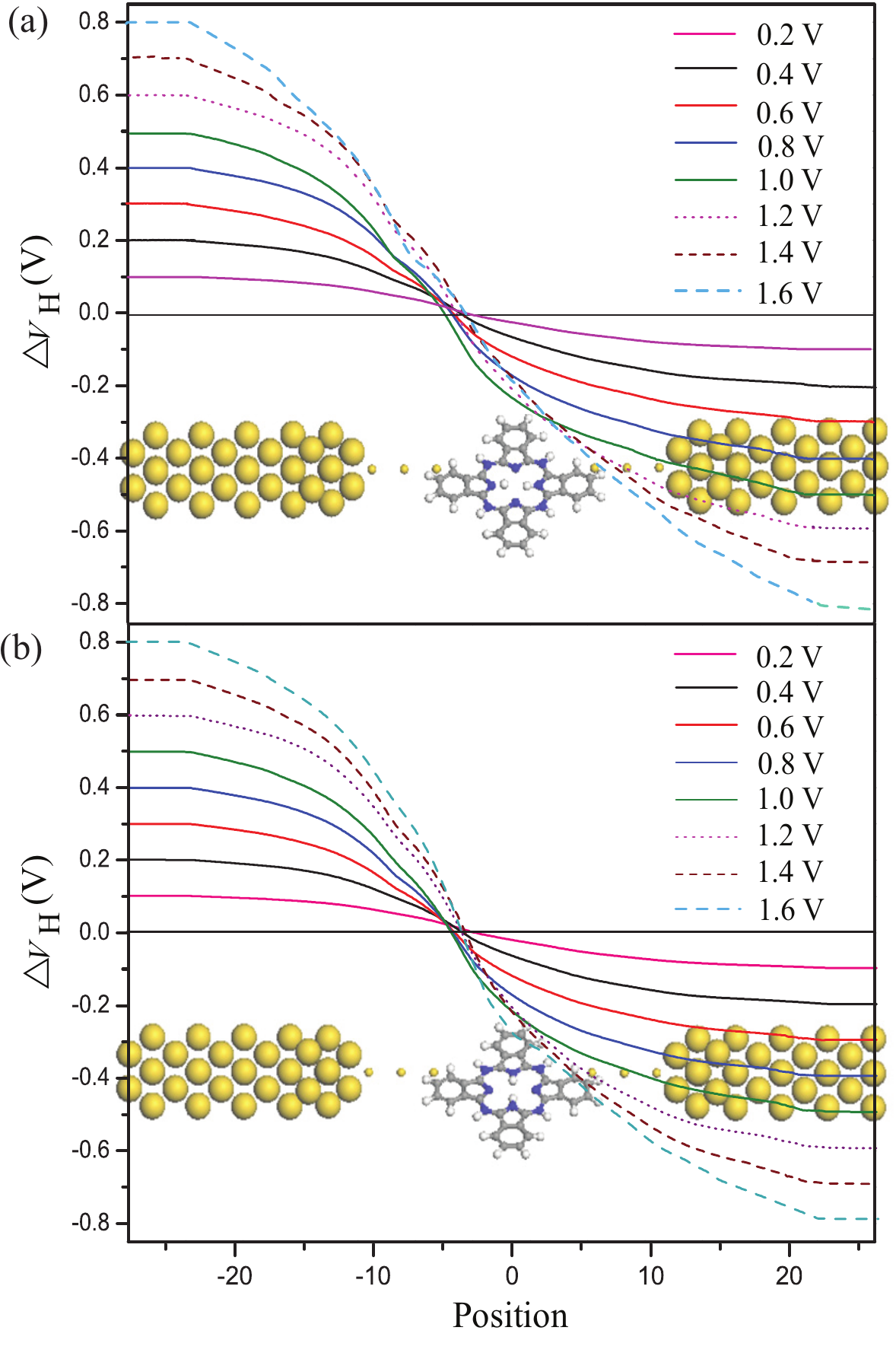}
\caption{(Color online) Electrostatic difference (Hartree) potential $\Delta V_H=V_H(V)-V_H(V=0)$,  showing the bias voltage drop over the junction in the (a) ON and (b) OFF states of Au(100)-Au$_3$-H$_{2}$Pc-Au$_3$-Au(100). For negative bias polarity one gets the mirror image of the bias voltage distribution.}
\label{Elstatpot_Au}
\end{figure}

Again, from the bias voltage dependence of the $T(E,V)$ transmission peaks in Fig. \ref{TEV_Au} we can draw a number of conclusions regarding charge transfer, level pinning, and bias voltage drop over the junction, similar to the case of cumulene wire leads:
For H$_{2}$Pc$_{\mbox{{\scriptsize h}}}$ (ON-spectrum), at low bias voltage in the range $0 \le V \le 0.4$, the current increases because the resonance peaks move distinctly slower than $\mu$ = \emph{E}$_{\mbox{{\scriptsize F}}}$-\emph{eV}/2, the bias window therefore including an increasing part of the line wings.
In the range $0.4 \le V \le 1$,  the transmission peaks move as fast as the chemical potential of the right lead, resulting in the plateau in the IVCs.
For $V>1$, something interesting happens: the intense resonance peaks shift back into the bias window and start following the chemical potential $\mu$ = \emph{E}$_{\mbox{{\scriptsize F}}}$+\emph{eV}/2 of the left lead, resulting in a large increase of the current in the range $1 \le V <1.4$,
Finally, for large bias voltage in the range of $1.4\le V\le1.6$, the width (and area) of the peak gets smaller, resulting in decreasing current and the appeareance of negative differential resistance (NDR). Also note that there are weak NDR effects in the $0.5\le V\le1.0$ region both for the Au$_2$ and Au$_3$ systems.
For H$_{2}$Pc$_{\mbox{{\scriptsize h}}}$ (OFF-spectrum), the pattern of IVCs is similar to that of the ON-spectrum, but different in magnitude.

The bias voltage drop over Au(100)-Au$_3$-H$_2$Pc-Au$_3$-Au(100) in the ON-state in  Fig. \ref{Elstatpot_Au}(a)  explains much of this picture:  like for cumulene leads (Fig. \ref{Elstatpot_cumulene}(a), the main bias voltage drop occurs at the left Au$_3$-H$_2$Pc contact, and most of the remaining bias voltage drop is distributed over the bulk of the H$_2$Pc molecule, the levels therefore basically following the chemical potential of the right lead with potential $\mu$ = \emph{E}$_{\mbox{{\scriptsize F}}}$-\emph{eV}/2 decreasing under applied bias.

However, there is in fact a clear difference: close inspection shows that the voltage drop in fact moves, with increasing bias across the left benzene ring toward the right lead, until around $V=1$ it jumps to the right H$_2$Pc-Au$_3$ interface, the chemical potential  $\mu$ = \emph{E}$_{\mbox{{\scriptsize F}}}$+\emph{eV}/2 of the left  lead now extending into the bulk of the H$_{2}$Pc$_{\mbox{{\scriptsize h}}}$ molecule. This is precisely where in the transmission spectrum in Fig. \ref{TEV_Au}(a) (top)  the transmisison peaks begin following the left lead with upgoing potential.
This means that the H$_{2}$Pc$_{\mbox{{\scriptsize h}}}$ molecule now begins to get negative charge, pinned by electron transfer from the left lead. The picture is also seen in the behaviour of the weak LUMO following the charging of the H$_{2}$Pc$_{\mbox{{\scriptsize h}}}$ molecule until it crosses the left upgoing leads, gets pinned, and then follows the left lead.

For the H$_{2}$Pc$_{\mbox{{\scriptsize v}}}$ OFF-case, the transmission spectrum is similar to that of the ON-spectrum, but different in magnitude, the transmission peaks being narrow and less intense, reflecting  much weaker coupling to the leads than in the ON-state. A particularly interesting feature is the different bias dependence of the leading transmission peaks for  $V>1$, essentially following the chemical potentials of different leads. This indicates that the HOMO and HOMO-1 orbitals at high bias are localized in quite different parts of the junction.

Finally, we conclude that the fundamental reason for the ON/OFF IVCs in both cumulene-H$_{2}$Pc$_{\mbox{{\scriptsize h}}}$-cumulene and Au-H$_{2}$Pc$_{\mbox{{\scriptsize h}}}$-Au is again the presence of two interacting, interfering, conduction channels around the Fermi level, being controlled by the tautomer switch. Although the bias-dependent charging of the molecular orbitals controls the positions of the transmission resonances relative to the bias voltage window, it is the interference effect that physically enhances the difference between the ON and OFF state transmissions. This is manifested in the Fano-type of line profiles characterizing the transmission peaks around the Fermi level.

%%%%%%%%%%%%%%%%%%%%%%%%%%%%%%%%%%%%%%%%%

\section{Comments on NDR effects}

The NDR effect in the case of Au-H$_{2}$Pc$_{\mbox{{\scriptsize h}}}$-Au around 1.6 V bias arises from the decreased intensity (area) of transmission peaks at high bias voltage, as shown in Fig. \ref{TEV_Au}. The reason is  reduced coupling between the Au-leads. Since the main conduction peaks follow the chemical potential of the left lead with upgoing potential, that part of the H$_2$Pc molecule should be well connected to the left lead. It is then reasonable to conclude (and supported by transmission eigenchannel analysis), that the coupling to the right lead is reduced.

Negative differential resistance is of great interest because NDR elements can be used for building bi-stable devices and highly functional components like latches, oscillators, memories  \cite{Reed2001,Husband2003} and logic XOR gates  \cite{Husband2003,Skoldberg2007a,Skoldberg2007b}. As a consequence, NDR has been at the focus of a number of recent experimental  \cite{Jangjian2009,Kang2010} and theoretical  \cite{Quek2007,Ribeiro2008,Pati2008} papers.

The NDR effect found by Quek \textit{et al.} \cite{Quek2007}  is explained in terms of gold electrodes supporting molecule-electrode contact driven NDR, where the coupling to the rather narrow bands of 1D Au chains stretched on the surface plays an important role. This looks similar to our case at high bias.

The NDR effect investigated by Riberio \textit{et al.} \cite{Ribeiro2008} deals with quantum transport properties of porphyrin-bridged p-n junctions with Si leads, and the NDR effect is controlled by the doping levels of the leads and the central transition metal atom of the porphyrin. The results are explained in terms of bias-induced on-off switching of resonant tunneling channels associated with specific molecular orbitals. This NDR can then be regarded as arising from bias-driven reduced molecule-lead overlap. This again looks similar to our case at high bias.

Finally, the NDR effect studied by Pati \textit{et al.} \cite{Pati2008} is explained in terms of reduced overlap between conduction links through the molecule itself. This is in contrast to our case where the reduced overlap appears at a contact to the molecule, the transmission peaks basically following one lead or the other. In the case of Ref.~\citenum{Pati2008}, 
the peaks do not move with bias, suggesting that the voltage drop over the molecule is distributed in a balanced way over and around the molecule.

From an IVC point of view there is no big difference between these cases. The differences are primarily visible in the $T(E,V)$ spectrum and reflect the details of the bias-dependence of the bias-voltage drop over the junctions, manifested in strongly non-linear IVCs. With more experience it might be possible to relate different types of NDR-features to different types of weak-link distributions and instabilities. It seems likely that voltage- or current- driven excitations and  instabilities should be sensitive to the distribution of strong fields (potential drops) over contacts or inside molecules.

Also note in Fig. \ref{IVC_Au} the appearance of weak NDR around 0.7 V bias in Au(001)-Au$_2$-H$_{2}$Pc-Au$_2$-Au(001), and very weak NDR around 0.9 V bias in  Au(001)-Au$_3$-H$_{2}$Pc-Au$_3$-Au(001). The latter case can in principle be understood from Fig. \ref{TEV_Au}a: the NDR (or plateau) seems to be connected with lineshape variation of the HOMO/HOMO-1 transmission peak structure. The Fano lineshapes suggest that interference effects again may play a role.

\section{Concluding remarks}

We have shown that the two tautomer configurations of phtalocyanine (H$_2$Pc) connected to conducting leads form junctions with widely different conductances, especially at low bias. With the H-H pair rotated (hydrogen tautomerization performed) by external means, the junction presents a single-molecular switch. This fact, together with the experimental demonstration of Liljeroth \textit{et al.} \cite{Liljeroth2007} that the switch works in principle in STM configuration (for vertical transport), makes H$_2$Pc an interesting candidate for a unimolecular electronic switch.

We find that hydrogen tautomerization affects the electronic state of H$_{2}$Pc by switching the order of the molecular orbital energies between HOMO and HOMO-1, causing a large change in the current. We specifically reveal the influence of molecular charging and molecule-electrode coupling on the transport mechanism. In particular, we show that the applied bias potential drop can shift from one metal-molecule contact to the other at intermediate bias, leading to pronounced features in the IVC. We also find that a two-channel interference effect is clearly manifested in pronounced Fano shapes characterizing the transmission peaks close to the Fermi level. This leads to the conclusion that the underlying reason for the ON/OFF IVCs in the studied configurations is the presence of two interacting, interfering, conduction channels around the Fermi level, being controlled by the tautomer switch. This interference effect is dramatic in the sense that it leads to a large ON/OFF ratio over a wide bias voltage range. 

Two final comments on the interference effect: (1) Since we expect the substrate to play a pronounced role stabilizing the two states of the molecule, the question arises how this will influence the interference, which is at the heart of the junction transport. Since the two nearly degenerate tautomer states have similar spacial localization, we expect the interaction with the substrate not to significantly change the level splitting. The levels could be shifted relative to the Fermi level, changing the conductivity, but that is not really the issue here: the ON/OFF switching contrast should remain.  (2) The calculation is done for frozen geometry, taking into account redistribution of electronic density on the molecule. The shift of the bias voltage drop in Fig. 7 in the range 1-1.4V is connected with the steep increase of conductivity in Fig. 2. Major electronic density redistribution can be expected to lead to conformational changes, and the large drop of the conductivity (NDR) in the1.4-1.6V range in Fig. 2 certainly needs to be investigated taking into account effects of atomic relaxation. But, then again, a free-molecule calculation may not be relevant: the atomic relaxation will be constrained by the stabilizing effect of the environment/substrate. For this reason, we expect the calculated NDR effect in Au-H2Pc-Au in Fig. 2 to be relevant for the molecule lying flat on a substrate.

We also conclude, that the D$_{2h}$ molecular symmetry of the H$_2$Pc with the HOMO and HOMO-1 occupying opposite `arms' with/without hydrogens is an essential component in creating the potential profile across the junction. It is important to use this symmetry in order to obtain different conductances for different H$_2$Pc configurations. The connection to the electrodes must not be to neighbouring arms (in which case the H-H pair rotation only flips the mirror symmetry of the junction): connecting neigboring arms produces an IVC that is the average of the ON- and OFF-states, losing the switching effect.
Coupling to the electrodes should also respect the symmetry by introducing minimal distortion. In our study H$_2$Pc is physisorbed at the gold electrodes; our attempts to chemisorb H$_2$Pc on gold via two thiol groups directly connected to the molecule only resulted in locking of the HOMO and HOMO-1 relative positions, thus preventing the switching. However, we exemplify with cumulene electrodes that it is possible to design linkers in the form of molecular ligands to interconnect H$_2$Pc to external leads. 

We demonstrate, by considering different electrode types, that the conduction of the junction is significantly affected by the choice of contacts. Both cumulene \cite{Prasongkit2010} and gold leads are highly conductive, however matching of the gold s-states to the H$_2$Pc conjugated $pi$-orbital introduces additional backscattering, yielding ~4 times less current at comparable voltages. In addition, the IVCs of junctions with gold leads exhibit pronounced negative differential resistance (NDR) at high bias voltage, as well as examples of weak NDR at intermediate bias. This, together with switching, makes H$_2$Pc an interesting candidate for a multifunctional molecular component \cite{Skoldberg2007a, Skoldberg2007b}. 

We realize, that the modelled system with the two closely spaced gold contacts is difficult to create in experiment. Nevertheless we believe that this difficulty is an additional good reason for preliminary theoretical assessment of the setup. Our investigation of H$_2$Pc is inspired by the need for switches and non-linear elements in molecular electronic circuits. However, the goal of this work has been strictly limited to investigating the transmission properties of the two tautomer states. We have not investigated the switching mechanism or stability of the OFF and ON states, or the influence of different substrates. 
It remains to investigate how the current through the junction may switch between OFF and ON states at elevated bias, including hysteretic behavior due to relaxation, current-driven fluctuations and thermal noise, as well as the influence of substrate surfaces on the conduction mechanisms.

\begin{acknowledgments}
J.P. is supported by the Royal Thai Government scholarship. A.G. and R.A. gratefully acknowledge financial support from Carl Tryggers Stiftelse f\"or Vetenskaplig Forskning and U3MEC, Uppsala. G.W. acknowledges support from the EU ICT-NANO projects NABAB and SINGLE. The calculations were performed at the high performance computing centers UPPMAX within the Swedish National Infrastructure for Computing.
\end{acknowledgments}

\end{document}